\documentclass[preprint,prd,aps,showpacs,showkeys,nofootinbib]{revtex4}
\usepackage{graphicx}
\usepackage{dcolumn}
\usepackage{bm}
\begin{document}

\preprint{ Submitted to ¡®Chinese Physics C¡¯}

\title{Oblique parameters in gauged baryon and lepton numbers with a $125\;{\rm GeV}$ Higgs}
\thanks{Supported by National Natural Science
Foundation of China (11247019) and Science and Technology Department of Liaoning (2012062)}

\author{Yang Xiu-Yi$^{1,2}$\footnote{yxyruxi@163.com}, Nie Jing$^{3,1}$}

\affiliation{$^1$ School of science,  University of Science and
Technology Liaoning, Anshan, Liaoning 114051, China \\$^2$ School of
Physics, Shandong University, Jinan, Shandong 250100, China\\$^3$Department of Physics, Dalian University of Technology, Dalian 116024, China}

\date{\today}

\begin{abstract}
In an extension of the standard model where baryon number and lepton number are local gauge symmetries,
we analyze the effect of corrections from exotic fermions and scalars on the oblique parameters $S,\;T,\;U$. Because a light neutral
Higgs $h_0$ with mass around $124-126{\rm GeV}$ strongly constrains the corresponding parameter space of this model , we also
investigate the gluon fusion process $gg\rightarrow h_0$ and two photon decay of lightest neutral Higgs
$h_0\rightarrow\gamma\gamma$ at the Large Hadron Collider.
\end{abstract}


\maketitle

\section{Introduction\label{sec1}}
\indent\indent
One of the main physics goals of the Large Hadron Collider (LHC) is to understand the origin of electroweak symmetry
breaking, and to search for the neutral Higgs boson predicted by the standard model (SM) and its various extensions.
Recently, ATLAS and CMS have reported significant excess events which are interpreted to be
most probably related to a neutral Higgs with mass $m_{h_0}\sim 124-126\;{\rm GeV}$. This implies that the Higgs mechanism
of breaking electroweak symmetry possibly has a solid experimental cornerstone.

The oblique parameters $S,\;T,\;U$~\cite{STU} are extracted from electroweak precision data (EWPD) observations
which probe the radiative corrections with sufficient accuracy. A light neutral Higgs with mass $124-126{\rm GeV}$ also affects
the theoretical evaluations of the oblique parameters $S,\;T,\;U$ through loop corrections to the gauge boson propagators,
which contain the neutral Higgs as a virtual field. In extensions of the SM, the corrections from exotic fields
to the gauge boson propagators can be expressed in terms of shifts of the parameters $S,\;T,\;U$~\cite{STU_NP}.

Broken baryon number (B) conservation can explain the origin of the matter-antimatter asymmetry in the Universe in a natural way.
Heavy majorana neutrinos contained in the seesaw mechanism can induce the tiny observed neutrino masses~\cite{seesaw}
to explain the results of neutrino oscillation experiments. Hence, lepton number (L) is also expected to be broken.
In~\cite{BL}, two extensions to the SM are examined, where $B$ and $L$ are spontaneously broken gauge
symmetries around the ${\rm TeV}$ scale, while ~\cite{BL_h} also investigates the predictions for the
Higgs mass and the Higgs decays in a supersymmetric model named BLMSSM, it is a
minimal supersymmetric extension of the SM (MSSM)with local gauged B and L.
Within the framework of the first extension of the SM with spontaneously broken  $B$ and $L$~\cite{BL},
we analyze the gluon fusion production and then decay into two photons
of the Higgs with mass $m_{h_0}\sim 124-126\;{\rm GeV}$. Additionally, we also investigate the corrections from
exotic fields of the oblique parameters $S,\;T,\;U$.

This paper is organized as follows. In Section~\ref{sec2}, we briefly summarize the main ingredients
of an extension of the SM where the baryon and lepton numbers are local symmetries, then present the mass squared
matrices for the neutral Higgs sector. Inspired by the new results from the ATLAS and CMS collaborations,
in Section~\ref{sec3} we study in great detail the Higgs production through gluon fusion followed by the decay of the Higgs boson
into two photons. We discuss the constraints on the parameter space
from the oblique parameters $S,\;T,\;U$ in Section~\ref{sec4}. Our conclusions are given
in Section~\ref{sec5}.

\section{An extension of the SM where baryon and lepton numbers are local gauge symmetries\label{sec2}}
\indent\indent
When baryon and lepton numbers are local gauge symmetries, one can write the gauge group as
$SU(3)_{_C}\otimes SU(2)_{_L}\otimes U(1)_{_Y}\otimes U(1)_{_B}\otimes U(1)_{_L}$.
In the first extension of the SM proposed in ~\cite{BL}, the exotic particles
include new quarks $Q_{_L}^\prime,\;u_{_R}^\prime,\;d_{_R}^\prime$, new leptons $l_{_L}^\prime,\;
\nu_{_R}^\prime,\;e_{_R}^\prime$ and three scalar singlets $S_{_B},\;S_{_L},\;S$ along with a scalar doublet $\phi$.
The Yukawa couplings are written as
\begin{eqnarray}
&&-{\cal L}_{_Y}=\Big\{\sum\limits_{I,J=1}^3\Big[\Big(Y_{_U}\Big)_{IJ}\bar{Q}_{_L}^I\tilde{H}u_{_R}^J
+\Big(Y_{_D}\Big)_{IJ}\bar{Q}_{_L}^IHd_{_R}^J\Big]+Y_{_U}^\prime\bar{Q}_{_L}^\prime Q_{_L}^{\prime c}S_{_B}
\nonumber\\
&&\hspace{1.5cm}
+\sum\limits_{I=1}^3\Big[\Big(Y_1\Big)_I\bar{Q}_{_L}^\prime\tilde{\phi}u_{_R}^J
+\Big(Y_2\Big)_I\bar{Q}_{_L}^I\phi d_{_R}^\prime\Big]+h.c.\Big\}
\nonumber\\
&&\hspace{1.5cm}
+\Big\{\sum\limits_{I,J=1}^3\Big[\Big(Y_{_\nu}\Big)_{IJ}\bar{L}_{_L}^I\tilde{H}\nu_{_R}^J
+\Big(Y_{_E}\Big)_{IJ}\bar{L}_{_L}^IHe_{_R}^J\Big]+Y_{_E}^\prime\bar{L}_{_L}^\prime L_{_L}^{\prime c} S_{_L}
\nonumber\\
&&\hspace{1.5cm}
+{1\over2}\sum\limits_{I,J=1}^3\Big(\lambda_a\Big)_{IJ}\bar{\nu}^{I,c}_{_R}S_{_L}^*\nu_{_R}^J
+\sum\limits_{I=1}^3\Big(\lambda_b\Big)_I\bar{\nu}_{_R}^{I,c}S_{_L}\nu_{_R}^\prime+h.c.\Big\}\;.
\label{Yukawa}
\end{eqnarray}
The scalar potential is generally given as follow:
\begin{eqnarray}
&&-{\cal L}_{_S}=m_{_H}^2H^\dagger H+m_{_\phi}^2\phi^\dagger\phi+m_{S_{_B}}^2S_{_B}^*S_{_B}
+m_{S_{_L}}^2S_{_L}^*S_{_L}+m_{_S}^2S^*S
\nonumber\\
&&\hspace{1.5cm}
+\lambda_{_{HH}}\Big(H^\dagger H\Big)\Big(H^\dagger H\Big)
+\lambda_{_{\phi\phi}}\Big(\phi^\dagger\phi\Big)\Big(\phi^\dagger\phi\Big)
+\lambda_{_{BB}}\Big(S_{_B}^*S_{_B}\Big)\Big(S_{_B}^*S_{_B}\Big)
\nonumber\\
&&\hspace{1.5cm}
+\lambda_{_{LL}}\Big(S_{_L}^*S_{_L}\Big)\Big(S_{_L}^*S_{_L}\Big)
+\lambda_{_{SS}}\Big(S^*S\Big)\Big(S^*S\Big)
+\lambda_{_{H\phi}}\Big(H^\dagger H\Big)\Big(\phi^\dagger\phi\Big)
\nonumber\\
&&\hspace{1.5cm}
+\lambda_{_{HB}}\Big(H^\dagger H\Big)\Big(S_{_B}^*S_{_B}\Big)
+\lambda_{_{HL}}\Big(H^\dagger H\Big)\Big(S_{_L}^*S_{_L}\Big)
+\lambda_{_{HS}}\Big(H^\dagger H\Big)\Big(S^*S\Big)
\nonumber\\
&&\hspace{1.5cm}
+\lambda_{_{\phi B}}\Big(\phi^\dagger\phi\Big)\Big(S_{_B}^*S_{_B}\Big)
+\lambda_{_{\phi L}}\Big(\phi^\dagger\phi\Big)\Big(S_{_L}^*S_{_L}\Big)
+\lambda_{_{\phi S}}\Big(\phi^\dagger\phi\Big)\Big(S^*S\Big)
\nonumber\\
&&\hspace{1.5cm}
+\lambda_{_{BL}}\Big(S_{_B}^*S_{_B}\Big)\Big(S_{_L}^*S_{_L}\Big)
+\lambda_{_{BS}}\Big(S_{_B}^*S_{_B}\Big)\Big(S^*S\Big)
+\lambda_{_{LS}}\Big(S_{_L}^*S_{_L}\Big)\Big(S^*S\Big)
\nonumber\\
&&\hspace{1.5cm}
+\lambda_{_{H\phi}}^\prime\Big(H^\dagger\phi\Big)\Big(\phi^\dagger H\Big)
+\Big[\mu_1\Big(H^\dagger\phi\Big)S+\mu_2S_{_B}^*S^2+h.c.\Big]\;.
\label{scalar-potential}
\end{eqnarray}
When the $SU(2)_L$ doublet $H$ and $SU(2)_L$ singlets $S_{_B},\;S_{_L}$ acquire the nonzero
vacuum expectation values (VEVs) $\upsilon,\;\upsilon_{_{B,L}}$,
\begin{eqnarray}
&&H=\left(\begin{array}{c}G^+\\{1\over\sqrt{2}}\Big(\upsilon+H_0+iG^0\Big)\end{array}\right)\;,
\nonumber\\
&&S_{_B}={1\over\sqrt{2}}\Big(\upsilon_{_B}+S_{_B}^0+iG_{_B}^0\Big)\;,
\nonumber\\
&&S_{_L}={1\over\sqrt{2}}\Big(\upsilon_{_L}+S_{_L}^0+iG_{_L}^0\Big)\;,
\label{VEVs}
\end{eqnarray}
the local gauge symmetry $SU(2)_{_L}\otimes U(1)_{_Y}\otimes U(1)_{_B}\otimes U(1)_{_L}$ is broken
down to the electromagnetic symmetry $U(1)_{_e}$, where $G^+,\;G^0,\;G_{_B}^0$ and $G_{_L}^0$ denote
the massless Goldstone bosons. Correspondingly, the mass terms for the neutral Higgs are formulated as
\begin{eqnarray}
&&-{\cal L}_{mass}^{H}={1\over2}\left(\begin{array}{ccc}H_0,\;&S_{_B}^0,\;&S_{_L}^0\end{array}\right)
m_{_{CPE}}^2\left(\begin{array}{c}H_0\\ S_{_B}^0 \\S_{_L}^0\end{array}\right)\;,
\label{mass-h0}
\end{eqnarray}
where the symmetric $3\times3$ mass squared matrix $m_{_{CPE}}^2$ is
\begin{eqnarray}
&&m_{_{CPE}}^2=\left(\begin{array}{ccc}2\lambda_{_{HH}}\upsilon^2&\lambda_{_{HB}}\upsilon\upsilon_{_B}&
\lambda_{_{HL}}\upsilon\upsilon_{_L}\\ \lambda_{_{HB}}\upsilon\upsilon_{_B}&2\lambda_{_{BB}}\upsilon_{_B}^2
&\lambda_{_{BL}}\upsilon_{_B}\upsilon_{_L}\\
\lambda_{_{HL}}\upsilon\upsilon_{_L}&\lambda_{_{BL}}\upsilon_{_B}\upsilon_{_L}&
2\lambda_{_{LL}}\upsilon_{_L}^2\end{array}\right)\;.
\label{mass-matrix1}
\end{eqnarray}
Through the orthogonal $3\times3$ transformation matrix $Z_{_{CPE}}$, the mass squared matrix $m_{_{CPE}}^2$
can be diagonalized as
\begin{eqnarray}
&&Z_{_{CPE}}^T m_{_{CPE}}^2Z_{_{CPE}}=\left(\begin{array}{ccc}m_{_{H_1^0}}^2,&m_{_{H_2^0}}^2,&
m_{_{H_3^0}}^2\end{array}\right)\;,
\label{diag1}
\end{eqnarray}
where $m_{_{H_1^0}}=m_{_{h_0}}\simeq125\;{\rm GeV}$.

In a similar way, we can write the $SU(2)_{_L}$ doublet $\phi$ and the $SU(2)_{_L}$ singlet $S$ as
\begin{eqnarray}
&&\phi=\left(\begin{array}{c}\phi^+\\{1\over\sqrt{2}}\Big(\phi_{_R}^0+i\phi_{_I}^0\Big)\end{array}\right)\;,
\nonumber\\
&&S={1\over\sqrt{2}}\Big(S_{_R}^0+iS_{_I}^0\Big)\;.
\label{scalar1}
\end{eqnarray}
As the local gauge symmetry $SU(2)_{_L}\otimes U(1)_{_Y}\otimes U(1)_{_B}\otimes U(1)_{_L}$ is broken
down to the electromagnetic symmetry $U(1)_{_e}$, the terms in square brackets in Eq. (\ref{scalar-potential})
induce mixing among the neutral scalar particles $\phi_{_R}^0,\;\phi_{_I}^0,\;S_{_R}^0,\;S_{_I}^0$,
and the mass terms are written as
\begin{eqnarray}
&&-{\cal L}_{mass}^{\Phi}={1\over2}\left(\begin{array}{cccc}\phi_{_R}^0,\;&S_{_R}^0,\;&\phi_{_I}^0,\;&S_{_I}^0\end{array}\right)
m_{_{CPM}}^2\left(\begin{array}{c}\phi_{_R}^0\\ S_{_R}^0\\\phi_{_I}^0 \\S_{_I}^0\end{array}\right)\;,
\label{mass-phi}
\end{eqnarray}
with the symmetric $4\times4$ mass squared matrix $m_{_{CPM}}^2$ being
\begin{eqnarray}
&&m_{_{CPM}}^2=\left(\begin{array}{cccc}m_{_\phi}^2&\sqrt{2}\upsilon\Re(\mu_1)&0&
-\sqrt{2}\upsilon\Im(\mu_1)\\
\sqrt{2}\upsilon\Re(\mu_1)&m_{_S}^2+2\sqrt{2}\upsilon_{_B}\Re(\mu_2)&0&-2\sqrt{2}\upsilon_{_B}\Im(\mu_2)\\
0&0&m_{_\phi}^2&-\sqrt{2}\upsilon\Re(\mu_1)\\
-\sqrt{2}\upsilon\Im(\mu_1)&-2\sqrt{2}\upsilon_{_B}\Im(\mu_2)&-\sqrt{2}\upsilon\Re(\mu_1)&
m_{_S}^2+2\sqrt{2}\upsilon_{_B}\Re(\mu_2)
\end{array}\right)\;.
\label{mass-matrix2}
\end{eqnarray}
We also diagonalize the mass squared matrix $m_{_{CPM}}^2$ through the $4\times4$ orthogonal rotation
$Z_{_{CPM}}$:
\begin{eqnarray}
&&Z_{_{CPM}}^T m_{_{CPM}}^2Z_{_{CPM}}=\left(\begin{array}{cccc}m_{_{\Phi_1^0}}^2,&m_{_{\Phi_2^0}}^2,&
m_{_{\Phi_3^0}}^2,&m_{_{\Phi_4^0}}^2\end{array}\right)\;.
\label{diag2}
\end{eqnarray}
Similarly, the mass for the charged scalar $\phi^\pm$ is expressed by
\begin{eqnarray}
&&m_{_{\phi^\pm}}^2={1\over2}m_{_\phi}^2-{1\over2}\lambda_{_{H\phi}}^\prime\upsilon^2\;.
\label{charged-phi}
\end{eqnarray}
Since the field $\phi$ does not get a nonzero VEV after the electroweak symmetry
is broken down, there is no mass mixing between the  exotic quarks and the SM quarks.

In the left-handed basis $(\nu_{_L}^I,\;\nu_{_L}^\prime,\;\nu_{_R}^{\prime c},\;\nu_{_R}^{I,c}),\;(I=1,\;2,\;3)$,
the mass matrix for neutrinos is given by the $8\times8$ matrix
\begin{eqnarray}
&&{\cal M}_{n}=\left(\begin{array}{cc}0_{3\times3}&\Big(M_{_D}\Big)_{3\times5}\\
\Big(M_{_D}^T\Big)_{5\times3}&\Big(M_{_N}\Big)_{5\times5}
\end{array}\right)\;.
\label{neutrino-matrix}
\end{eqnarray}
Here, the $3\times5$ matrix $M_{_D}$ is written as
\begin{eqnarray}
&&M_{_D}=\left(\begin{array}{ccc}0_{3\times2},&{\upsilon\over\sqrt{2}}\Big(Y_\nu^*\Big)_{3\times3}\end{array}\right)\;,
\label{neutrino-matrixD}
\end{eqnarray}
and the $5\times5$ matrix $M_{_N}$ is
\begin{eqnarray}
&&M_{_N}=\left(\begin{array}{ccc}0,&{\upsilon\over\sqrt{2}}Y_\nu^{\prime*},&0_{1\times3}\\
{\upsilon\over\sqrt{2}}Y_\nu^{\prime*},&0,&{\upsilon_{_L}\over\sqrt{2}}\Big(\lambda_b^*\Big)_{1\times3}\\
0_{3\times1},&{\upsilon_{_L}\over\sqrt{2}}\Big(\lambda_b^\dagger\Big)_{3\times1},&
{\upsilon_{_L}\over\sqrt{2}}\Big(\lambda_a^\dagger\Big)_{3\times3}
\end{array}\right)\;,
\label{neutrino-matrixN}
\end{eqnarray}
Integrating the heavy freedoms out, we get the following mass matrix for three light neutrinos:
\begin{eqnarray}
&&{\cal M}_\nu=-M_{_D}M_{_N}^{-1}M_{_D}^T\;,
\label{light-neutrino-matrix}
\end{eqnarray}
which is diagonalized by the Pontecorvo-Maki-Nakagawa-Sakata matrix $U_{_{PMNS}}$
\begin{eqnarray}
&&U_{_{PMNS}}^T{\cal M}_\nu U_{_{PMNS}}={\it diag}({m_{_{\nu_1}},\;m_{_{\nu_2}},\;m_{_{\nu_3}}})\;.
\label{dia-neutrino-matrix1}
\end{eqnarray}
Meanwhile, the Majorana mass matrix $M_{_N}$ is similarly diagonalized by a $5\times5$ matrix
$Z_{_N}$
\begin{eqnarray}
&&Z_{_N}^TM_{_N}Z_{_N}={\it diag}({m_{_{N_1}},\;m_{_{N_2}},\;m_{_{N_3}}},\;m_{_{N_4}},\;m_{_{N_5}})\;.
\label{dia-neutrino-matrix2}
\end{eqnarray}

\section{The $gg\rightarrow h_0\rightarrow\gamma\gamma$ process in gauged baryon and lepton numbers\label{sec3}}
\indent\indent
At the LHC, the Higgs is produced chiefly through gluon fusion.  In the SM, the leading order (LO) contributions originate
from the one-loop diagram, which involves virtual top quarks. The cross section for this process is known to
the next-to-next-to-leading order (NNLO)~\cite{NNLO}, which can enhance the LO result by 80-100\%. Furthermore, any new particle
which couples strongly with the Higgs can significantly modify this cross section. In the extension of the SM considered here,
the LO decay width for the process $h_0\rightarrow gg$ is given by (see ~\cite{Gamma1} and references therein)
\begin{eqnarray}
&&\Gamma_{_{NP}}(h_0\rightarrow gg)={G_{_F}\alpha_s^2m_{_{h_0}}^3|(Z_{_{CPE}})_{_{11}}|^2\over64\sqrt{2}\pi^3}
\Big|A_{1/2}(x_t)+A_{1/2}(x_{t^\prime})+A_{1/2}(x_{b^\prime})\Big|^2\;,
\label{hgg}
\end{eqnarray}
where $x_a=m_{_{h_0}}^2/(4m_a^2),\;a=t,\;t^\prime,\;b^\prime$, and the loop function $A_{1/2}$ is defined in the Appendix.

The Higgs to diphoton decay is also obtained from loop diagrams. The LO contributions are derived from the
one-loop diagrams containing virtual charged gauge bosons $W^\pm$ or virtual top quarks
in the SM. In this model, the additional charged scalar $\phi^\pm$ and exotic fermions $t^\prime,\;b^\prime,\;\tau^\prime$
contribute corrections to the decay width of the Higgs to diphoton at LO. The corresponding expression is written as
\begin{eqnarray}
&&\Gamma_{_{NP}}(h_0\rightarrow\gamma\gamma)={G_{_F}\alpha^2m_{_{h_0}}^3\over128\sqrt{2}\pi^3}
\Big|(Z_{_{CPE}})_{_{11}}\Big({4\over3}A_{1/2}(x_t)+{4\over3}A_{1/2}(x_{t^\prime})+{1\over3}A_{1/2}(x_{b^\prime})
\nonumber\\
&&\hspace{2.7cm}
+A_{1/2}(x_{\tau^\prime})+A_1(x_{_{\rm W}})\Big)+{8m_{_{\rm W}}^2s_{_{\rm W}}^2\over e^2m_{_{\phi^\pm}}^2}
\Big(\lambda_{_{H\phi}}(Z_{_{CPE}})_{_{11}}+{\upsilon_{_B}\over\upsilon}\lambda_{_{\phi B}}(Z_{_{CPE}})_{_{21}}
\nonumber\\
&&\hspace{2.7cm}
+{\upsilon_{_L}\over\upsilon}\lambda_{_{\phi L}}(Z_{_{CPE}})_{_{31}}\Big)A_0(x_{\phi^\pm})\Big|^2\;,
\label{hpp}
\end{eqnarray}
where the concrete expressions for the loop functions $A_0,\;A_1$ are given in the Appendix.

The Higgs discovery from both the ATLAS and CMS experiments have  observed an excess
in Higgs production and decay into the diphoton channel which is a factor $1.4\sim2$ times larger than
the SM expectations. The observed signal for the diphoton channels is quantified by the ratio
\begin{eqnarray}
&&R_{\gamma\gamma}={\Gamma_{_{NP}}(h_0\rightarrow gg)\Gamma_{_{NP}}(h_0\rightarrow\gamma\gamma)\over
\Gamma_{_{SM}}(h_0\rightarrow gg)\Gamma_{_{SM}}(h_0\rightarrow\gamma\gamma)}\;,
\label{signal}
\end{eqnarray}
where we assume that all exotic fields are heavier than the lightest Higgs $h_0$.
The current value of this ratio is as follows~\cite{CMS,ATLAS}:
\begin{eqnarray}
&&{\rm ATLAS}:\;\;R_{\gamma\gamma}=1.90\pm0.5\;,
\nonumber\\
&&{\rm CMS}:\;\;R_{\gamma\gamma}=1.56\pm0.43\;,
\nonumber\\
&&{\rm ATLAS+CMS}:\;\;R_{\gamma\gamma}=1.71\pm0.33\;.
\label{signal-exp}
\end{eqnarray}
Note that the combination of the ATLAS and CMS results is taken from ~\cite{CMS-ATLAS}.

\section{Corrections to the oblique parameters\label{sec4}}
\indent\indent
A common approach to constrain physics beyond the SM is to use global electroweak
fitting through the oblique parameters $S,\;T,\;U$~\cite{STU}. In the SM, electroweak
precision tests imply a relationship between $m_{_{h_0}}$ and $m_{_{\rm Z}}$.
In the model considered here, the electroweak precision tests also strongly constrain the mass spectrum and
relevant couplings.

Here, we adopt the definitions of the oblique parameters $S,\;T,\;U$ given in ~\cite{STU,He-Su}:
\begin{eqnarray}
&&S=16\pi\Big\{{\Pi_{33}(m_{_{\rm Z}}^2)-\Pi_{33}(0)\over m_{_{\rm Z}}^2}
-{\Pi_{3Q}(m_{_{\rm Z}}^2)-\Pi_{3Q}(0)\over m_{_{\rm Z}}^2}\Big\}\;,\nonumber\\
&&T=4\pi{\Pi_{11}(0)-\Pi_{33}(0)\over m_{_{\rm Z}}^2s_{_{\rm W}}^2c_{_{\rm W}}^2}\;,\nonumber\\
&&U=16\pi\Big\{{\Pi_{11}(m_{_{\rm Z}}^2)-\Pi_{11}(0)\over m_{_{\rm Z}}^2}
-{\Pi_{33}(m_{_{\rm Z}}^2)-\Pi_{33}(0)\over m_{_{\rm Z}}^2}\Big\}\;,
\label{STU-def}
\end{eqnarray}
where $s_{_{\rm W}}=\sin\theta_{_{\rm W}} and \;c_{_{\rm W}}=\cos\theta_{_{\rm W}}$ with the Weinberg
angle $\theta_{_{\rm W}}$ defined at the energy scale $\mu=m_{_{\rm Z}}$. In the above definitions,
$\Pi_{11}$ and $\Pi_{33}$ are the vacuum polarizations of isospin currents, and $\Pi_{3Q}$ is
the vacuum polarization of one isospin and one hypercharge current.

By comparing the measurable electroweak observables with the theoretical predictions,
one finds the fitted values\cite{STU_exp}
\begin{eqnarray}
&&\Delta S=S-S_{_{\rm SM}}=0.04\pm0.10\;,\nonumber\\
&&\Delta T=T-T_{_{\rm SM}}=0.05\pm0.11\;,\nonumber\\
&&\Delta U=U-U_{_{\rm SM}}=0.08\pm0.11\;.
\label{STU-exp}
\end{eqnarray}

As mentioned above, there is no mass mixing between the exotic quarks and the SM quarks.
The corresponding corrections to the oblique parameters from exotic quarks are
\begin{eqnarray}
&&\Delta S_{Q^\prime}={1\over\pi}\Big\{\int_0^1dxx(1-x)\ln{m_{b^\prime}^2-x(1-x)m_{_{\rm Z}}^2\over m_{t^\prime}^2-x(1-x)m_{_{\rm Z}}^2}
-{3m_{t^\prime}^2\over2m_{_{\rm Z}}^2}\int_0^1dx\ln{m_{t^\prime}^2-x(1-x)m_{_{\rm Z}}^2\over m_{t^\prime}^2}
\nonumber\\
&&\hspace{1.5cm}
-{3m_{b^\prime}^2\over2m_{_{\rm Z}}^2}\int_0^1dx\ln{m_{b^\prime}^2-x(1-x)m_{_{\rm Z}}^2\over m_{b^\prime}^2}\Big\}\;,
\nonumber\\
&&\Delta T_{Q^\prime}=-{3\over4\pi s_{_{\rm W}}^2c_{_{\rm W}}^2}\Big\{{m_{t^\prime}^2\over m_{_{\rm Z}}^2}\int_0^1dx
x\ln{xm_{t^\prime}^2+(1-x)m_{_{b^\prime}}^2\over m_{t^\prime}^2}
+{m_{b^\prime}^2\over m_{_{\rm Z}}^2}\int_0^1dx(1-x)\ln{xm_{t^\prime}^2+(1-x)m_{_{b^\prime}}^2\over m_{b^\prime}^2}\Big\}\;,
\nonumber\\
&&\Delta U_{Q^\prime}={1\over\pi}\Big\{3\int_0^1dxx(1-x)\ln{[xm_{t^\prime}^2+(1-x)m_{b^\prime}^2-x(1-x)m_{_{\rm Z}}^2]^2\over
[m_{t^\prime}^2-x(1-x)m_{_{\rm Z}}^2][m_{b^\prime}^2-x(1-x)m_{_{\rm Z}}^2]}
\nonumber\\
&&\hspace{1.5cm}
-3\int_0^1dx\Big(x{m_{t^\prime}^2\over m_{_{\rm Z}}^2}+(1-x){m_{b^\prime}^2\over m_{_{\rm Z}}^2}\Big)
\ln{xm_{t^\prime}^2+(1-x)m_{b^\prime}^2-x(1-x)m_{_{\rm Z}}^2\over xm_{t^\prime}^2+(1-x)m_{b^\prime}^2}
\nonumber\\
&&\hspace{1.5cm}
-{3m_{t^\prime}^2\over2m_{_{\rm Z}}^2}\int_0^1dx\ln{m_{t^\prime}^2-x(1-x)m_{_{\rm Z}}^2\over m_{t^\prime}^2}
-{3m_{b^\prime}^2\over2m_{_{\rm Z}}^2}\int_0^1dx\ln{m_{b^\prime}^2-x(1-x)m_{_{\rm Z}}^2\over m_{b^\prime}^2}\Big\}\;.
\label{oblique-quarks}
\end{eqnarray}
Here, $m_{b^\prime} and \;m_{t^\prime}$ denote the masses of the charged $-1/3$ exotic quark $b^\prime$ and the charged $2/3$
exotic quark $t^\prime$, respectively.

In a similar way, there is no mass mixing between the exotic charged leptons and the SM leptons.
Ignoring the tiny mixing between the left-handed neutrinos and heavy majorana neutrinos,
we write the corrections to the oblique parameters from exotic leptons as
\begin{eqnarray}
&&\Delta S_{L^\prime}={1\over\pi}\sum\limits_{i,j=1}^5\Big(Z_{_N}\Big)_{1i}\Big(Z_{_N}^\dagger\Big)_{i1}
\Big(Z_{_N}\Big)_{1j}\Big(Z_{_N}^\dagger\Big)_{j1}\Big\{-{1\over2}\int_0^1dx\Big(x{m_{_{N_i}}^2\over m_{_{\rm Z}}^2}
+(1-x){m_{_{N_j}}^2\over m_{_{\rm Z}}^2}\Big)
\nonumber\\
&&\hspace{1.5cm}\times
\ln{xm_{_{N_i}}^2+(1-x)m_{_{N_j}}^2-x(1-x)m_{_{\rm Z}}^2\over xm_{_{N_i}}^2+(1-x)m_{_{N_j}}^2}
-{m_{\tau^\prime}^2\over2m_{_{\rm Z}}^2}\int_0^1dx\ln{m_{\tau^\prime}^2-x(1-x)m_{_{\rm Z}}^2\over m_{\tau^\prime}^2}
\nonumber\\
&&\hspace{1.5cm}
+\int_0^1dxx(1-x)\ln{xm_{_{N_i}}^2+(1-x)m_{_{N_j}}^2-x(1-x)m_{_{\rm Z}}^2\over
m_{\tau^\prime}^2-x(1-x)m_{_{\rm Z}}^2}\Big\}\;,
\nonumber\\
&&\Delta T_{L^\prime}=-{1\over4\pi s_{_{\rm W}}^2c_{_{\rm W}}^2}\sum\limits_{i,j=1}^5\Big(Z_{_N}\Big)_{1i}\Big(Z_{_N}^\dagger\Big)_{i1}
\Big(Z_{_N}\Big)_{1j}\Big(Z_{_N}^\dagger\Big)_{j1}\Big\{{m_{_{N_i}}^2\over m_{_{\rm Z}}^2}\int_0^1dx
\ln{xm_{_{N_i}}^2+(1-x)m_{\tau^\prime}^2\over xm_{_{N_i}}^2+(1-x)m_{_{N_j}}^2}
\nonumber\\
&&\hspace{1.5cm}
+{m_{\tau^\prime}^2\over m_{_{\rm Z}}^2}\int_0^1dx(1-x)\ln{xm_{_{N_i}}^2+(1-x)m_{\tau^\prime}^2\over m_{\tau^\prime}^2}\Big\}\;,
\nonumber\\
&&\Delta U_{L^\prime}={1\over\pi}\sum\limits_{i,j=1}^5\Big(Z_{_N}\Big)_{1i}\Big(Z_{_N}^\dagger\Big)_{i1}
\Big(Z_{_N}\Big)_{1j}\Big(Z_{_N}^\dagger\Big)_{j1}\Big\{
{m_{\tau^\prime}^2\over2m_{_{\rm Z}}^2}\int_0^1dx\ln{m_{\tau^\prime}^2-x(1-x)m_{_{\rm Z}}^2\over m_{\tau^\prime}^2}
\nonumber\\
&&\hspace{1.5cm}
+\int_0^1dxx(1-x)\ln{[xm_{_{N_i}}^2+(1-x)m_{\tau^\prime}^2-x(1-x)m_{_{\rm Z}}^2]^2\over
(xm_{_{N_i}}^2+(1-x)m_{_{N_j}}^2-x(1-x)m_{_{\rm Z}}^2)(m_{\tau^\prime}^2-x(1-x)m_{_{\rm Z}}^2)}
\nonumber\\
&&\hspace{1.5cm}
-\int_0^1dx\Big(x{m_{_{N_i}}^2\over m_{_{\rm Z}}^2}+(1-x){m_{\tau^\prime}^2\over m_{_{\rm Z}}^2}\Big)
\ln{xm_{_{N_i}}^2+(1-x)m_{\tau^\prime}^2-x(1-x)m_{_{\rm Z}}^2\over xm_{_{N_i}}^2+(1-x)m_{\tau^\prime}^2}
\nonumber\\
&&\hspace{1.5cm}
+{1\over2}\int_0^1dx\Big(x{m_{_{N_i}}^2\over m_{_{\rm Z}}^2}+(1-x){m_{_{N_j}}^2\over m_{_{\rm Z}}^2}\Big)
\ln{xm_{_{N_i}}^2+(1-x)m_{_{N_j}}^2-x(1-x)m_{_{\rm Z}}^2\over xm_{_{N_i}}^2+(1-x)m_{_{N_j}}^2}\Big\}\;.
\label{oblique-leptons}
\end{eqnarray}
Here, the $5\times5$ unitary matrix $Z_{_N}$ is the mixing matrix for heavy majorana neutrinos,
$m_{_{N_i}}\;(i=1,\;2,\;\cdots,\;5)$ are the corresponding masses of the heavy neutrinos, and $m_{\tau^\prime}$
is the mass of the charged exotic lepton $\tau^\prime$.

As the radiative corrections to the self energy of gauge bosons originate from three CP-even Higgs
$(h_0,\;H_2^0,\;H_3^0)$, the corresponding contributions to the oblique parameters are given by
\begin{eqnarray}
&&\Delta S_{H}={1\over\pi}\sum\limits_{i=1}^3(Z_{_{CPE}})_{1i}^2\Big\{
{1\over2}\int_0^1dxx(1-x)\ln{x^2m_{_{\rm Z}}^2+(1-x)m_{_{H_i^0}}^2\over m_{_{\rm Z}}^2}
\nonumber\\
&&\hspace{1.5cm}
+\int_0^1dx\Big(1-{x\over2}-(1-x){m_{_{H_i^0}}^2\over2m_{_{\rm Z}}^2}\Big)
\ln{x^2m_{_{\rm Z}}^2+(1-x)m_{_{H_i^0}}^2\over xm_{_{\rm Z}}^2+(1-x)m_{_{H_i^0}}^2}\Big\}\;,
\nonumber\\
&&\Delta T_{H}={1\over4\pi s_{_{\rm W}}^2c_{_{\rm W}}^2}
\sum\limits_{i=1}^3(Z_{_{CPE}})_{1i}^2\Big\{
-\int_0^1dx\Big(1-{x\over2}-(1-x){m_{_{H_i^0}}^2\over2m_{_{\rm Z}}^2}\Big)
\nonumber\\
&&\hspace{1.5cm}\times
\ln{xm_{_{\rm Z}}^2+(1-x)m_{_{H_i^0}}^2\over m_{_{\rm Z}}^2}
\nonumber\\
&&\hspace{1.5cm}
+\int_0^1dx\Big((1-{x\over2})c_{_{\rm W}}^2-(1-x){m_{_{H_i^0}}^2\over2m_{_{\rm Z}}^2}\Big)
\ln{xm_{_{\rm W}}^2+(1-x)m_{_{H_i^0}}^2\over m_{_{\rm Z}}^2}\Big\}\;,
\nonumber\\
&&\Delta U_{H}={1\over\pi}\sum\limits_{i=1}^3(Z_{_{CPE}})_{1i}^2\Big\{
{1\over2}\int_0^1dx\Big(x^2+(1-x){m_{_{H_i^0}}^2\over m_{_{\rm Z}}^2}\Big)
\nonumber\\
&&\hspace{1.5cm}\times
\ln{x^2m_{_{\rm Z}}^2+(1-x)m_{_{H_i^0}}^2\over m_{_{\rm Z}}^2}
-\int_0^1dx\ln{x^2m_{_{\rm Z}}^2+(1-x)m_{_{H_i^0}}^2\over xm_{_{\rm Z}}^2+(1-x)m_{_{H_i^0}}^2}
\nonumber\\
&&\hspace{1.5cm}
-{1\over2}\int_0^1dx\Big(x+(1-x){m_{_{H_i^0}}^2\over m_{_{\rm Z}}^2}\Big)
\ln{xm_{_{\rm Z}}^2+(1-x)m_{_{H_i^0}}^2\over m_{_{\rm Z}}^2}
\nonumber\\
&&\hspace{1.5cm}
+\int_0^1dx\Big((1-{x\over2})c_{_{\rm W}}^2-(1-x){m_{_{H_i^0}}^2\over2m_{_{\rm Z}}^2}\Big)
\ln{xm_{_{\rm W}}^2+(1-x)m_{_{H_i^0}}^2-x(1-x)m_{_{\rm Z}}^2\over xm_{_{\rm W}}^2+(1-x)m_{_{H_i^0}}^2}
\nonumber\\
&&\hspace{1.5cm}
+{1\over2}\int_0^1dxx(1-x)\ln{xm_{_{\rm W}}^2+(1-x)m_{_{H_i^0}}^2-x(1-x)m_{_{\rm Z}}^2
\over m_{_{\rm Z}}^2}\Big\}\;.
\label{oblique-Higgs}
\end{eqnarray}
Here we adopt the notation $H_1^0$ to represent the lightest neutral Higgs $h_0$. In addition,
the contributions from $\Phi_i^0$ and $\phi^\pm$ to the oblique parameters are formulated
as follows
\begin{eqnarray}
&&\Delta S_{\phi}={1\over2\pi}\Big\{\sum\limits_{i,j}^4(Z_{_{CPM}})_{1i}^2(Z_{_{CPM}})_{3j}^2
\Big[-\int_0^1dx\Big(x{m_{_{\Phi_i^0}}^2\over m_{_{\rm Z}}^2}+(1-x){m_{_{\Phi_j^0}}^2\over m_{_{\rm Z}}^2}\Big)
\nonumber\\
&&\hspace{1.5cm}\times
\ln{xm_{_{\Phi_i^0}}^2+(1-x)m_{_{\Phi_j^0}}^2-x(1-x)m_{_{\rm Z}}^2\over
xm_{_{\Phi_i^0}}^2+(1-x)m_{_{\Phi_j^0}}^2}
\nonumber\\
&&\hspace{1.5cm}
-\int_0^1dxx(1-x)\ln{xm_{_{\Phi_i^0}}^2+(1-x)m_{_{\Phi_j^0}}^2-x(1-x)m_{_{\rm Z}}^2
\over m_{_{\phi^\pm}}^2-x(1-x)m_{_{\rm Z}}^2}\Big]
\nonumber\\
&&\hspace{1.5cm}
+{m_{_{\phi^\pm}}^2\over m_{_{\rm Z}}^2}\int_0^1dx\ln{m_{_{\phi^\pm}}^2-x(1-x)m_{_{\rm Z}}^2
\over m_{_{\phi^\pm}}^2}\Big\}\;,
\nonumber\\
&&\Delta T_{\phi}={1\over8\pi s_{_{\rm W}}^2c_{_{\rm W}}^2}\Big\{
-\sum\limits_{i=1}^4\Big[(Z_{_{CPM}})_{1i}^2+(Z_{_{CPM}})_{3i}^2\Big]
\nonumber\\
&&\hspace{1.5cm}\times
\Big[{m_{_{\phi^\pm}}^2\over m_{_{\rm Z}}^2}\int_0^1dxx\ln{xm_{_{\phi^\pm}}^2+(1-x)m_{_{\Phi_i^0}}^2
\over m_{_{\phi^\pm}}^2}
+{m_{_{\Phi_i^0}}^2\over m_{_{\rm Z}}^2}\int_0^1dxx\ln{xm_{_{\Phi_i^0}}^2+(1-x)m_{_{\phi^\pm}}^2
\over m_{_{\Phi_i^0}}^2}\Big]
\nonumber\\
&&\hspace{1.5cm}
+\sum\limits_{i,j}^4(Z_{_{CPM}})_{1i}^2(Z_{_{CPM}})_{3j}^2
\int_0^1dx\Big(x{m_{_{\Phi_i^0}}^2\over m_{_{\rm Z}}^2}+(1-x){m_{_{\Phi_j^0}}^2\over m_{_{\rm Z}}^2}\Big)
\nonumber\\
&&\hspace{1.5cm}\times
\ln{xm_{_{\Phi_i^0}}^2+(1-x)m_{_{\Phi_j^0}}^2\over m_{_{\Phi_i^0}}^2}\Big\}\;,
\nonumber\\
&&\Delta U_{\phi}={1\over2\pi}\Big\{
-\sum\limits_{i=1}^4\Big[(Z_{_{CPM}})_{1i}^2+(Z_{_{CPM}})_{3i}^2\Big]
\Big[\int_0^1dx\Big(x{m_{_{\phi^\pm}}^2\over m_{_{\rm Z}}^2}+(1-x){m_{_{\Phi_i^0}}^2\over m_{_{\rm Z}}^2}\Big)
\nonumber\\
&&\hspace{1.5cm}\times
\ln{xm_{_{\phi^\pm}}^2+(1-x)m_{_{\Phi_i^0}}^2-x(1-x)m_{_{\rm Z}}^2\over xm_{_{\phi^\pm}}^2+(1-x)m_{_{\Phi_i^0}}^2}
\nonumber\\
&&\hspace{1.5cm}
+{1\over2}\int_0^1dxx(1-x)\ln{xm_{_{\phi^\pm}}^2+(1-x)m_{_{\Phi_i^0}}^2-x(1-x)m_{_{\rm Z}}^2
\over m_{_{\phi^\pm}}^2-x(1-x)m_{_{\rm Z}}^2}\Big]
\nonumber\\
&&\hspace{1.5cm}
+\sum\limits_{i,j}^4(Z_{_{CPM}})_{1i}^2(Z_{_{CPM}})_{3j}^2
\Big[\int_0^1dx\Big(x{m_{_{\Phi_i^0}}^2\over m_{_{\rm Z}}^2}+(1-x){m_{_{\Phi_j^0}}^2\over m_{_{\rm Z}}^2}\Big)
\nonumber\\
&&\hspace{1.5cm}\times
\ln{xm_{_{\Phi_i^0}}^2+(1-x)m_{_{\Phi_j^0}}^2-x(1-x)m_{_{\rm Z}}^2\over xm_{_{\Phi_i^0}}^2+(1-x)m_{_{\Phi_j^0}}^2}
\nonumber\\
&&\hspace{1.5cm}
+\int_0^1dxx(1-x)\ln{xm_{_{\Phi_i^0}}^2+(1-x)m_{_{\Phi_j^0}}^2-x(1-x)m_{_{\rm Z}}^2\over
xm_{_{\Phi_i^0}}^2+(1-x)m_{_{\phi^\pm}}^2-x(1-x)m_{_{\rm Z}}^2}
\nonumber\\
&&\hspace{1.5cm}
+\int_0^1dxx(1-x)\ln{xm_{_{\Phi_i^0}}^2+(1-x)m_{_{\Phi_j^0}}^2-x(1-x)m_{_{\rm Z}}^2\over
xm_{_{\phi^\pm}}^2+(1-x)m_{_{\Phi_j^0}}^2-x(1-x)m_{_{\rm Z}}^2}\Big]
\nonumber\\
&&\hspace{1.5cm}
+{m_{_{\phi^\pm}}^2\over m_{_{\rm Z}}^2}\int_0^1dx\ln{m_{_{\phi^\pm}}^2-x(1-x)m_{_{\rm Z}}^2
\over m_{_{\phi^\pm}}^2}\Big\}\;.
\label{oblique-Phi}
\end{eqnarray}

\section{Numerical analysis\label{sec5}}
\indent\indent
As mentioned above, the most stringent constraint on the parameter space is that the
$3\times3$ mass squared matrix in Eq.(\ref{mass-matrix1}) should produce the lightest eigenvector
with a mass $m_{_{h_0}}=125\;{\rm GeV}$.

 In order to make the final results consistent with this condition,
we require the self coupling of the Higgs doublet to satisfy
\begin{eqnarray}
&&\lambda_{_{HH}}={A\over B}\;,
\label{Higgs-mass1}
\end{eqnarray}
where
\begin{eqnarray}
&&A=m_{_{h_0}}^6-2\Big[\lambda_{_{BB}}\upsilon_{_B}^2+\lambda_{_{LL}}\upsilon_{_L}^2\Big]m_{_{h_0}}^4
+\Big[(4\lambda_{_{BB}}\lambda_{_{LL}}-\lambda_{_{BL}}^2)\upsilon_{_B}^2\upsilon_{_L}^2
-\lambda_{_{HB}}^2\upsilon^2\upsilon_{_B}^2
\nonumber\\
&&\hspace{1.0cm}
-\lambda_{_{HL}}^2\upsilon^2\upsilon_{_L}^2\Big]m_{_{h_0}}^2
+2\Big[\lambda_{_{BB}}\lambda_{_{HL}}^2+\lambda_{_{LL}}\lambda_{_{HB}}^2
-\lambda_{_{HB}}\lambda_{_{HL}}\lambda_{_{BL}}\Big]\upsilon^2\upsilon_{_B}^2\upsilon_{_L}^2\;,
\nonumber\\
&&B=2\upsilon^2\Big[m_{_{h_0}}^4-2(\lambda_{_{BB}}\upsilon_{_B}^2+\lambda_{_{LL}}\upsilon_{_L}^2)m_{_{h_0}}^2
+(4\lambda_{_{BB}}\lambda_{_{LL}}-\lambda_{_{BL}}^2)\upsilon_{_B}^2\upsilon_{_L}^2\Big]\;.
\label{Higgs-mass2}
\end{eqnarray}

The present experimental lower bounds on the fourth generation charged lepton $\tau'$, up-type and down-type
quarks $t'$ and $b'$ at $95$ $\%$ C.L. are $m_{\tau'}>100.8\;{\rm GeV}\;$, $m_{t'}>420\;{\rm GeV}\;$ and
$m_{b'}>372\;{\rm GeV}\;$, respectively. The fourth generation quarks  $t'$ and $b'$ acquire nonzero masses $m_{t'}=m_{b'}=\frac{Y_{Q}'}{\sqrt{2}}\upsilon_{_B}$
when local $U(1)_{B}$ symmetry is broken. In addition, the charged leptons of the fourth generation $\tau'$ obtains nonzero masses $m_{\tau'}=\frac{Y_{E}'}{\sqrt{2}}\upsilon_{_L}$
when local $U(1)_{L}$ symmetry is broken.

However, there are too many free parameters in the model considered here. In our numerical analysis, we adopt the assumption on the parameter space
\begin{eqnarray}
&&m_{t^\prime}=m_{b^\prime}=m_{\tau^\prime}=m_{_F}\;,\nonumber\\
&&\lambda_{_{BB}}=\lambda_{_{LL}}=0.5\;,\;\;\lambda_{_{HL}}=\lambda_{_{BL}}=\lambda_{_{HB}}=\lambda_{_{NP}}\;,
\nonumber\\
&&\lambda_{_{\phi H}}=\lambda_{_{\phi B}}=\lambda_{_{\phi L}}=\lambda_{_{NP}}^\prime\;,
\label{assumption1}
\end{eqnarray}
to decrease the number of free parameters in the concerned model.
Furthermore, we assume $\upsilon\ll\upsilon_{_{B,L}}$,
$\Big(\lambda_a\Big)_{3\times3}={\it diag}(\lambda_a,\;\lambda_a,\;\lambda_a)$, and choose the hierarchical assumption
on Yukawa couplings $\Big|(Y_\nu)_{IJ}\Big|\ll|Y_\nu^\prime|\sim\lambda_a\sim\lambda_{b_{I}},\;\;(I,\;J=1,\;2,\;3)$
to obtain our final results. Applying the assumptions above, we obtain the majorana mass for
the lightest exotic neutrino $N_1$ to be
\begin{eqnarray}
&&m_{_{N_1}}\simeq{\upsilon^2\over\sqrt{2}\upsilon_{_L}}{\lambda_a|Y_\nu^\prime|\over\lambda_b^2}\;,
\label{majorana-neutrino-mass1}
\end{eqnarray}
with $\lambda_b^2=\lambda_{_{b_1}}^2+\lambda_{_{b_2}}^2+\lambda_{_{b_3}}^2$.
Of course, we need this mass to be greater than $m_{_{\rm Z}}/2$ to be consistent with the measured
$Z$-boson decay width. The masses of other heavy majorana neutrinos are
\begin{eqnarray}
&&m_{_{N_i}}\simeq\Big({\upsilon_{_L}\over\sqrt{2}}\lambda_a,\;{\upsilon_{_L}\over\sqrt{2}}\lambda_a,\;
{\upsilon_{_L}\over2\sqrt{2}}(\Delta-\lambda_a),
\;{\upsilon_{_L}\over2\sqrt{2}}(\Delta+\lambda_a)\Big)\;,
\label{majorana-neutrino-mass2}
\end{eqnarray}
for $i=2,\;3,\;4,\;5$ and $\Delta=\sqrt{\lambda_a^2+4\lambda_b^2}$.

Correspondingly, the $5\times5$ mixing matrix $Z_N$ is approximated as
\begin{eqnarray}
&&Z_N\simeq\left(\begin{array}{ccccc}1,&-{\lambda_aY_\nu^*\upsilon\over\lambda_b^2\upsilon_{_L}},&
{\lambda_{_{b_1}}Y_\nu^*\upsilon\over\lambda_b^2\upsilon_{_L}},&
{\lambda_{_{b_2}}Y_\nu^*\upsilon\over\lambda_b^2\upsilon_{_L}},&
{\lambda_{_{b_3}}Y_\nu^*\upsilon\over\lambda_b^2\upsilon_{_L}}\\
{\lambda_aY_\nu^*\upsilon\over\lambda_b^2\upsilon_{_L}},&
0,&0,&-i\sqrt{\Delta+\lambda_a\over2\Delta},&\sqrt{\Delta-\lambda_a\over2\Delta}\\
-{\lambda_{_{b_1}}Y_\nu^*\upsilon\over\lambda_b^2\upsilon_{_L}},&
-{\lambda_{_{b_3}}\over\sqrt{\lambda_{_{b_1}}^2+\lambda_{_{b_3}}^2}},&
-{\lambda_{_{b_2}}\over\sqrt{\lambda_{_{b_1}}^2+\lambda_{_{b_2}}^2}},&
{i\sqrt{2}\lambda_{_{b_1}}\over\sqrt{\Delta^2+\lambda_a\Delta}},&
{\sqrt{2}\lambda_{_{b_1}}\over\sqrt{\Delta^2-\lambda_a\Delta}}\\
-{\lambda_{_{b_2}}Y_\nu^*\upsilon\over\lambda_b^2\upsilon_{_L}},&
0,&0,&{i\sqrt{2}\lambda_{_{b_2}}\over\sqrt{\Delta^2+\lambda_a\Delta}},&
{\sqrt{2}\lambda_{_{b_2}}\over\sqrt{\Delta^2-\lambda_a\Delta}}\\
-{\lambda_{_{b_3}}Y_\nu^*\upsilon\over\lambda_b^2\upsilon_{_L}},&
{\lambda_{_{b_1}}\over\sqrt{\lambda_{_{b_1}}^2+\lambda_{_{b_3}}^2}},
&{\lambda_{_{b_1}}\over\sqrt{\lambda_{_{b_1}}^2+\lambda_{_{b_2}}^2}},&
{i\sqrt{2}\lambda_{_{b_3}}\over\sqrt{\Delta^2+\lambda_a\Delta}},&
{\sqrt{2}\lambda_{_{b_3}}\over\sqrt{\Delta^2-\lambda_a\Delta}}
\end{array}\right)\;.
\label{majorana-neutrino-mixing1}
\end{eqnarray}

For the relevant parameters in the SM, we take~\cite{PDG}
\begin{eqnarray}
&&\alpha_s(m_{_{\rm Z}})=0.118\;,\;\;\alpha(m_{_{\rm Z}})=1/128\;,
\nonumber\\
&&s_{_{\rm W}}^2(m_{_{\rm Z}})=0.23\;,\;\;m_t=174.2\;{\rm GeV}\;,\;\;
m_{_{\rm W}}=80.4\;{\rm GeV}\;.
\label{PDG-SM}
\end{eqnarray}
\begin{figure}[b]
\setlength{\unitlength}{1mm}
\centering
\includegraphics[width=3.15in]{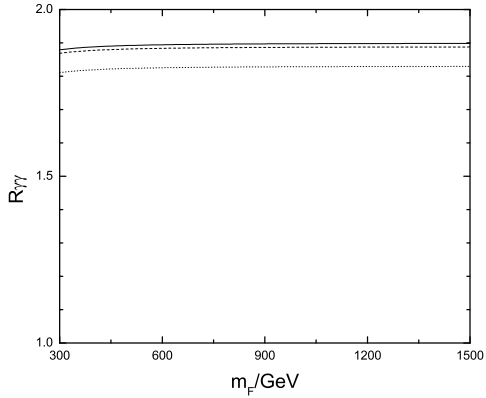}
\vspace{-1cm}
\caption[]{Variation of $R_{\gamma\gamma}$ with the mass scale of exotic fermions $m_{_F}$ when
$m_{\phi^\pm}=500{\rm GeV},\;\lambda_{NP}=0.5,\;\lambda_{NP}^\prime=-0.5$. The dotted line represents
$\upsilon_{_B}=\upsilon_{_L}=500\;{\rm GeV}$, the dashed line represents $\upsilon_{_B}=\upsilon_{_L}=1000\;{\rm GeV}$,
and the solid line represents $\upsilon_{_B}=\upsilon_{_L}=1500\;{\rm GeV}$.}
\label{fig1}
\end{figure}
\subsection{Numerical results of $R_{\gamma\gamma}$}
\begin{figure}[b]
\setlength{\unitlength}{1mm}
\centering
\includegraphics[width=3.15in,height=6.6in]{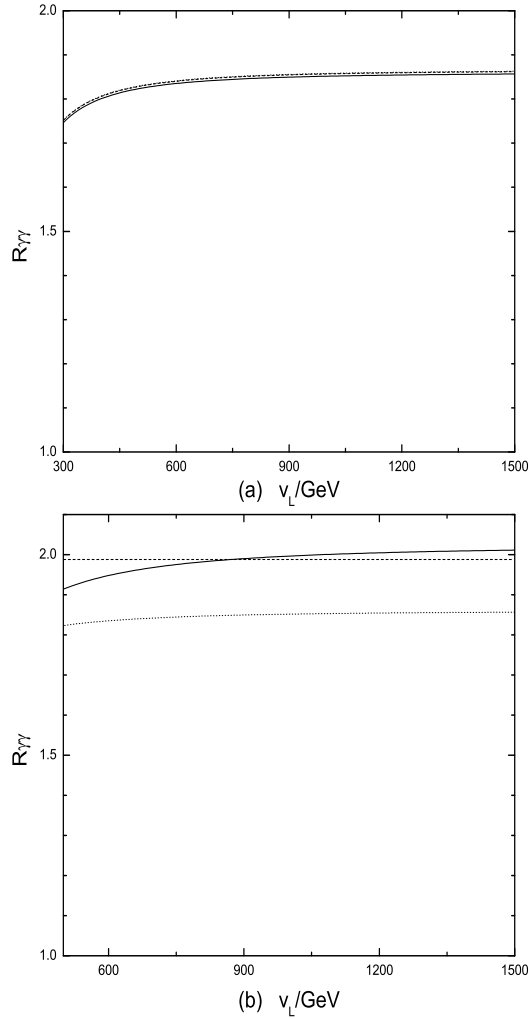}
\vspace{-2cm}
\caption[]{Variation of $R_{\gamma\gamma}$ with the VEV $\upsilon_{_L}$ when $m_{\phi^\pm}
=\upsilon_{_B}=500{\rm GeV},\;\lambda_{NP}^\prime=-0.5$. In (a), $\lambda_{NP}=0.5$,
the dotted line corresponds to $m_{_F}=500{\rm GeV}$, the dashed line corresponds to $m_{_F}=550\;{\rm GeV}$,
and the solid line corresponds to $m_{_F}=600\;{\rm GeV}$. In (b), $m_{_F}=500{\rm GeV}$,
the dotted line corresponds to $\lambda_{NP}=0.5$, the dashed line corresponds to $\lambda_{NP}=0$,
and the solid line corresponds to $\lambda_{NP}=-0.5$.}
\label{fig2}
\end{figure}
Under our assumptions on the parameter space, the theoretical prediction of $R_{\gamma\gamma}$
depends on six parameters in the model: $m_{_F},\;m_{\phi^\pm},\;\lambda_{NP},\;\lambda_{NP}^\prime,\;
\upsilon_{_B}$ and $\upsilon_{_L}$. Taking $m_{\phi^\pm}=500{\rm GeV},\;\lambda_{NP}=0.5 and \;\lambda_{NP}^\prime=-0.5$,
we plot the variation of $R_{\gamma\gamma}$ with the mass scalar of exotic fermions $M_{_F}$, as shown in Fig.\ref{fig1}.
The dotted line corresponds to $\upsilon_{_B}=\upsilon_{_L}=500\;{\rm GeV}$, the dashed line corresponds
to $\upsilon_{_B}=\upsilon_{_L}=1000\;{\rm GeV}$, and the solid line corresponds to
$\upsilon_{_B}=\upsilon_{_L}=1500\;{\rm GeV}$.
In general, the ratio $R_{\gamma\gamma}$
depends very weakly on the mass scale $m_{_F}$, and the value of $R_{\gamma\gamma}$
is about $1.8\sim1.9$ when $500\;{\rm GeV}\le\upsilon_{_B}=\upsilon_{_L}\le1500\;{\rm GeV}$.
\begin{figure}[b]
\setlength{\unitlength}{1mm}
\centering
\includegraphics[width=3.15in,height=6.0in]{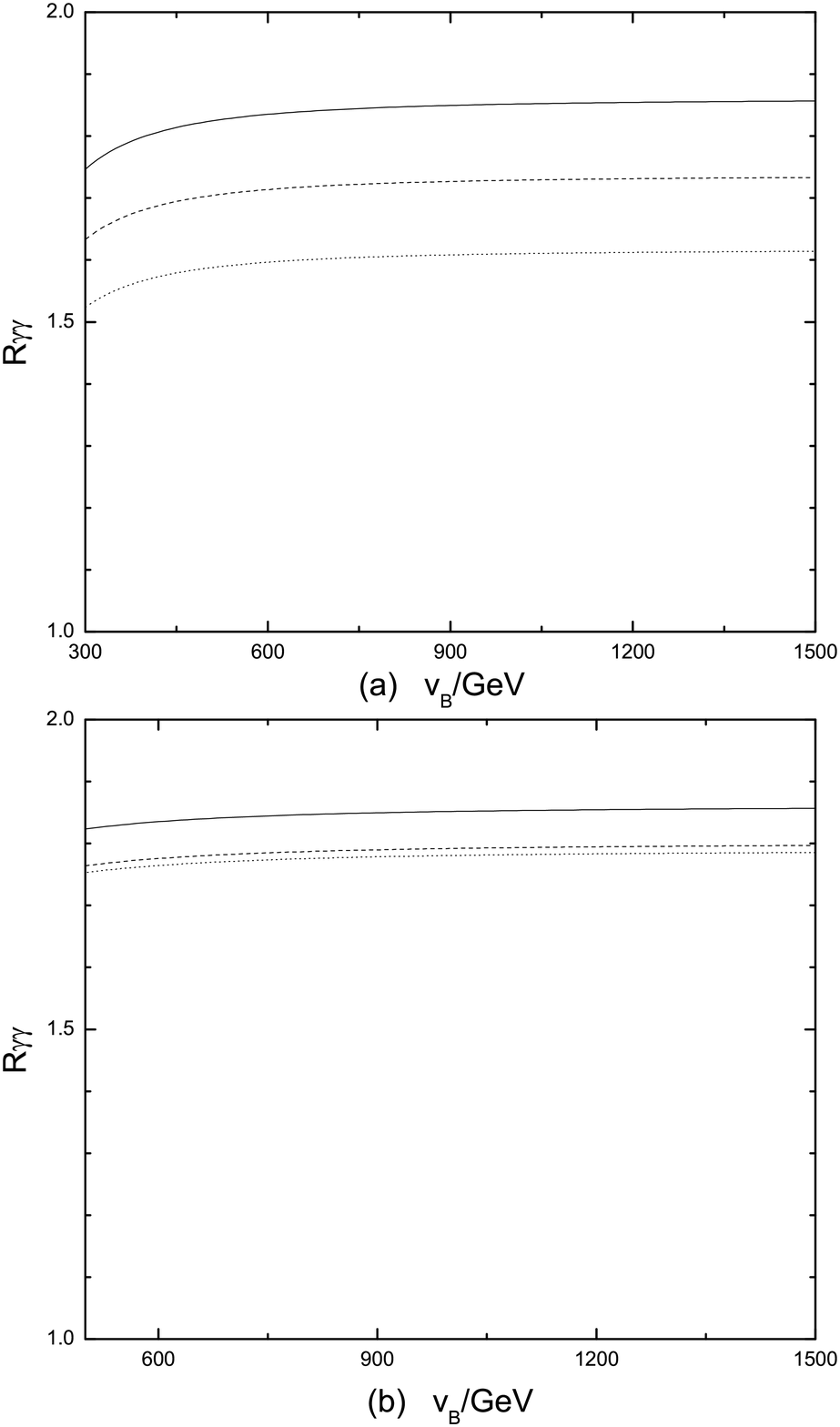}
\vspace{-2cm}
\caption[]{Variation of $R_{\gamma\gamma}$ with the VEV $\upsilon_{_B}$ when $m_{_F}
=\upsilon_{_L}=500{\rm GeV},\;\lambda_{NP}=0.5$ for: (a) $m_{\phi^\pm}=500\;{\rm GeV}$, where
the dotted line corresponds to $\lambda_{NP}^\prime=0.5$, the dashed line corresponds to $\lambda_{NP}^\prime=0$,
and the solid line corresponds to $\lambda_{NP}^\prime=-0.5$;  (b)$\lambda_{NP}^\prime=-0.5$, where
the dotted line corresponds to $m_{\phi^\pm}=1500\;{\rm GeV}$, the dashed line corresponds to $m_{\phi^\pm}=1000\;{\rm GeV}$,
and the solid line corresponds to $m_{\phi^\pm}=500\;{\rm GeV}$.}
\label{fig3}
\end{figure}

In Fig.~\ref{fig2}(a), we plot the variation of $R_{\gamma\gamma}$ with the VEV $\upsilon_{_L}$
when $m_{\phi^\pm}=\upsilon_{_B}=500\;{\rm GeV},\;\lambda_{NP}^\prime=-0.5$ and $\lambda_{NP}=0.5$.
The dotted line corresponds to $m_{_F}=500{\rm GeV}$, the dashed line corresponds to $m_{_F}=550\;{\rm GeV}$,
and the solid line corresponds to $m_{_F}=600\;{\rm GeV}$. The dependence of
$R_{\gamma\gamma}$ on $\upsilon_{_L}$ is relatively sensitive for $\upsilon_{_L}\le600\;{\rm GeV}$,
and is weak for $\upsilon_{_L}>600\;{\rm GeV}$. Since the dependence of  $R_{\gamma\gamma}$ on $m_{_F}$ and $\upsilon_{_B}$ is very weak, the three lines almost coincide with each other.
In Fig.~\ref{fig2}(b), we plot the variation of $R_{\gamma\gamma}$ with the VEV $\upsilon_{_L}$
when $m_{_F}=m_{\phi^\pm}=\upsilon_{_B}=500{\rm GeV},\;\lambda_{NP}^\prime=-0.5$. The dotted line corresponds to $\lambda_{NP}=0.5$, the dashed line corresponds to $\lambda_{NP}=0$, and the solid line corresponds to $\lambda_{NP}=-0.5$. Generally, there is a weak dependence of the ratio
$R_{\gamma\gamma}$ on $\upsilon_{_L}$.

In Fig.~\ref{fig3}(a), we show the variation of $R_{\gamma\gamma}$ with the VEV $\upsilon_{_B}$
when $m_{_F}=\upsilon_{_L}=500\;{\rm GeV},\;\lambda_{NP}=0.5$.
The dotted line corresponds to $\lambda_{NP}^\prime=0.5$, the dashed line corresponds to $\lambda_{NP}^\prime=0$,
and the solid line corresponds to $\lambda_{NP}^\prime=-0.5$. The dependence of
$R_{\gamma\gamma}$ on $\upsilon_{_B}$ is relatively sensitive for $\upsilon_{_B}\le600\;{\rm GeV}$,
and is weak for $\upsilon_{_B}>600\;{\rm GeV}$.
In Fig.~\ref{fig3}(b), we show the variation of $R_{\gamma\gamma}$ with $\upsilon_{_B}$
when $m_{_F}=\upsilon_{_L}=500\;{\rm GeV},\;\lambda_{NP}^\prime=-0.5$ and $\lambda_{NP}=0.5$.
The dotted line corresponds to $m_{\phi^\pm}=1500{\rm GeV}$, the dashed line corresponds to $m_{\phi^\pm}=1000\;{\rm GeV}$,
and the solid line corresponds to $m_{\phi^\pm}=500\;{\rm GeV}$. Generally, there is a very weak dependence of the ratio
$R_{\gamma\gamma}$ on $\upsilon_{_B}$.
\begin{figure}[h]
\setlength{\unitlength}{1mm}
\centering
\includegraphics[width=3.15in]{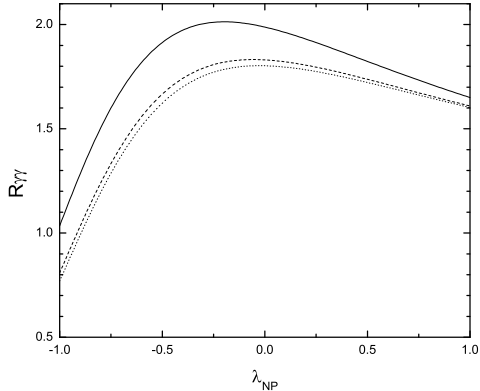}
\vspace{-1cm}
\caption[]{Variation of $R_{\gamma\gamma}$ with $\lambda_{NP}$ when
$\upsilon_{_B}=\upsilon_{_L}=500\;{\rm GeV},\;\lambda_{NP}^\prime=-0.5$, where the dotted line represents
$m_{_F}=500\;{\rm GeV},m_{\phi^\pm}=1500\;{\rm GeV}$, the dashed line represents $m_{_F}=550\;{\rm GeV},m_{\phi^\pm}=1000\;{\rm GeV}$,
and the solid line represents $m_{_F}=m_{\phi^\pm}=500\;{\rm GeV}$.}
\label{fig4}
\end{figure}
\begin{figure}[h]
\setlength{\unitlength}{1mm}
\centering
\includegraphics[width=3.15in]{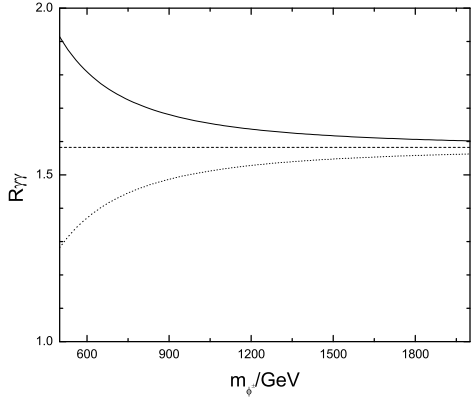}
\vspace{-1cm}
\caption[]{Variation of $R_{\gamma\gamma}$ with $m_{\phi^\pm}$ when
$m_{_F}=\upsilon_{_B}=\upsilon_{_L}=500\;{\rm GeV},\;\lambda_{NP}=-0.5$, where the dotted line represents
$\lambda_{NP}^\prime=0.5$, the dashed line represents $\lambda_{NP}^\prime=0$,
and the solid line represents $\lambda_{NP}^\prime=-0.5$.}
\label{fig5}
\end{figure}

Choosing $\upsilon_{_B}=\upsilon_{_L}=500{\rm GeV}, \;\lambda_{NP}^\prime=-0.5$,
Fig.\ref{fig4} presents the variation of the ratio $R_{\gamma\gamma}$ with $\lambda_{NP}$.
The dotted line represents $m_{_F}=500\;{\rm GeV},m_{\phi^\pm}=1500\;{\rm GeV}$, the dashed line
represents $m_{_F}=550\;{\rm GeV},m_{\phi^\pm}=1000\;{\rm GeV}$, and the solid line represents
$m_{_F}=m_{\phi^\pm}=500\;{\rm GeV}$. As $\Lambda_{NP}$ increases,
$R_{\gamma\gamma}$ changes drastically and can easily coincide with the present experimental data,
as $-0.5\le\lambda_{_{NP}}\le1.0$.
Choosing $m_{_F}=\upsilon_{_B}=\upsilon_{_L}=500\;{\rm GeV}, and \;\lambda_{NP}=-0.5$,
Fig.\ref{fig5} shows the ratio $R_{\gamma\gamma}$ versus $m_{\phi^\pm}$.
The dotted line represents  $\lambda_{NP}^\prime=0.5$, the dashed line represents $\lambda_{NP}^\prime=0$,
and the solid line represents $\lambda_{NP}^\prime=-0.5$. For $\lambda_{NP}^\prime=0$, there is a slight dependence of
$R_{\gamma\gamma}$ on the mass $m_{\phi^\pm}$. When $\lambda_{NP}^\prime=\pm0.5$,
$R_{\gamma\gamma}$ decreases steeply as $m_{\phi^\pm}$ increases.

\indent\indent
\begin{figure}[b]
\setlength{\unitlength}{1mm}
\centering
\includegraphics[width=3.15in]{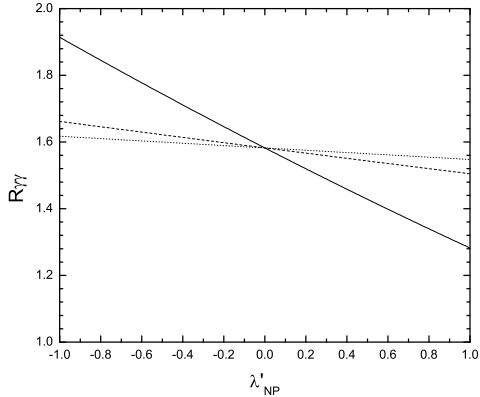}
\vspace{-1cm}
\caption[]{Variation of $R_{\gamma\gamma}$ with $\lambda_{NP}^\prime$ when
$m_{_F}=\upsilon_{_B}=\upsilon_{_L}=500\;{\rm GeV},\;\lambda_{NP}=-0.5$, where the dotted line represents
$m_{\phi^\pm}=1500\;{\rm GeV}$, the dashed line represents $m_{\phi^\pm}=1000\;{\rm GeV}$,
and the solid line represents $m_{\phi^\pm}=500\;{\rm GeV}$.}
\label{fig6}
\end{figure}
In Fig.~\ref{fig6}, we plot the variation of the ratio $R_{\gamma\gamma}$ with $\lambda_{NP}^\prime$
when $m_{_F}=\upsilon_{_B}=\upsilon_{_L}=500\;{\rm GeV} and \;\lambda_{NP}=-0.5$. The dotted line represents
$m_{\phi^\pm}=1500\;{\rm GeV}$, the dashed line represents $m_{\phi^\pm}=1000\;{\rm GeV}$,
and the solid line represents $m_{\phi^\pm}=500\;{\rm GeV}$. The dependence of
$R_{\gamma\gamma}$ on $\lambda_{NP}^\prime$ is strong when $m_{\phi^\pm}=500\;{\rm GeV}$
but weaker for higher values of $m_{\phi^\pm}$.

Generally,  the ratio $R_{\gamma\gamma}$  depends strongly on the parameters $\lambda_{NP},\;\lambda_{NP}^\prime$
and $m_{\phi^\pm}$, and depends weakly on $\upsilon_{_{B}},\upsilon_{_L}$ and $m_{_F}$. These numerical results can be
reasonably explained  from Eq.(\ref{hgg}) and Eq.(\ref{hpp}), where $\lambda_{NP}$ affects theoretical predictions
of $R_{\gamma\gamma}$ through the $3\times3$ mixing matrix $Z_{_{CPE}}$, while $\lambda_{NP}^\prime and m_{\phi^\pm}$
affect theoretical predictions of $R_{\gamma\gamma}$ through the last term in Eq.(\ref{hpp}).

The important point is that the parameters $\lambda_a,\;\lambda_{b_i}\;(i=1,\;2,\;3)$ do not
affect the theoretical predictions of $R_{\gamma\gamma}$ since there is no correction to the decay widths
of $h_0\rightarrow\gamma\gamma$ and $h_0\rightarrow gg$ from the neutrino sector at one-loop level.
Similarly, the parameters $M_{_S},\;\mu_1,\;\mu_2$ also do not affect theoretical evaluations of
$R_{\gamma\gamma}$, as there is no one-loop correction to the decay widths of $h_0\rightarrow\gamma\gamma$
and $h_0\rightarrow gg$ from virtual $\Phi_i^0\;(i=1,\;2,\;3,\;4)$.

\subsection{The constraints on parameter space from oblique corrections}
\indent\indent
The heavy neutrinos contribute one-loop radiative corrections to the self energies of $ZZ,\;W^\pm W^\mp$
in this model. This results in the theoretical values of the $S,\;T,\;U$ parameters depending on
$\lambda_a,\;\lambda_{b_i}\;(i=1,\;2,\;3)$ here. Furthermore, the theoretical values of the $S,\;T,\;U$ parameters also depend on
$m_{_S},\;\mu_1,\;\mu_2$ through the virtual $\phi^\pm,\;\Phi_{i}^0,\;(i=1,\;2,\;3,\;4)$ radiative corrections to
the self energies of $ZZ,\;W^\pm W^\mp$ at one-loop level. So far, fitting $S,\;T,\;U$ within $3\sigma$ deviation indicates
\begin{eqnarray}
&&-0.26\le\Delta S\le0.34\;,
\nonumber\\
&&-0.28\le\Delta T\le0.38\;,
\nonumber\\
&&-0.25\le\Delta U\le0.41\;.
\label{fitting1}
\end{eqnarray}
In order to obtain theoretical values of
$S,\;T,\;U$ which satisfy present experimental data, we adopt the additional assumptions here:
\begin{eqnarray}
&&m_{_{N_1}}\simeq{\upsilon^2\over\sqrt{2}\upsilon_{_L}}{\lambda_a|Y_\nu^\prime|\over\lambda_b^2}=m_{_F}\;,
\nonumber\\
&&\lambda_{b_1}=\lambda_{b_2}=\lambda_{b_3}={1\over\sqrt{3}}\lambda_b\;,
\nonumber\\
&&\upsilon_{_B}=\upsilon_{_L}=m_{_S}=m_{_F}=500\;{\rm GeV}\;,
\nonumber\\
&&|\mu_1|=20\;{\rm GeV},\;|\mu_2|=200\;{\rm GeV}\;,
\nonumber\\
&&\lambda_a=\lambda_b=0.6,\;\lambda_{_{BB}}=\lambda_{_{LL}}=0.5,\;
\lambda_{_{NP}}=\lambda_{_{NP}}^\prime=0.01\;.
\label{assumption1}
\end{eqnarray}
\begin{figure}[h]
\setlength{\unitlength}{1mm}
\centering
\includegraphics[width=3.15in]{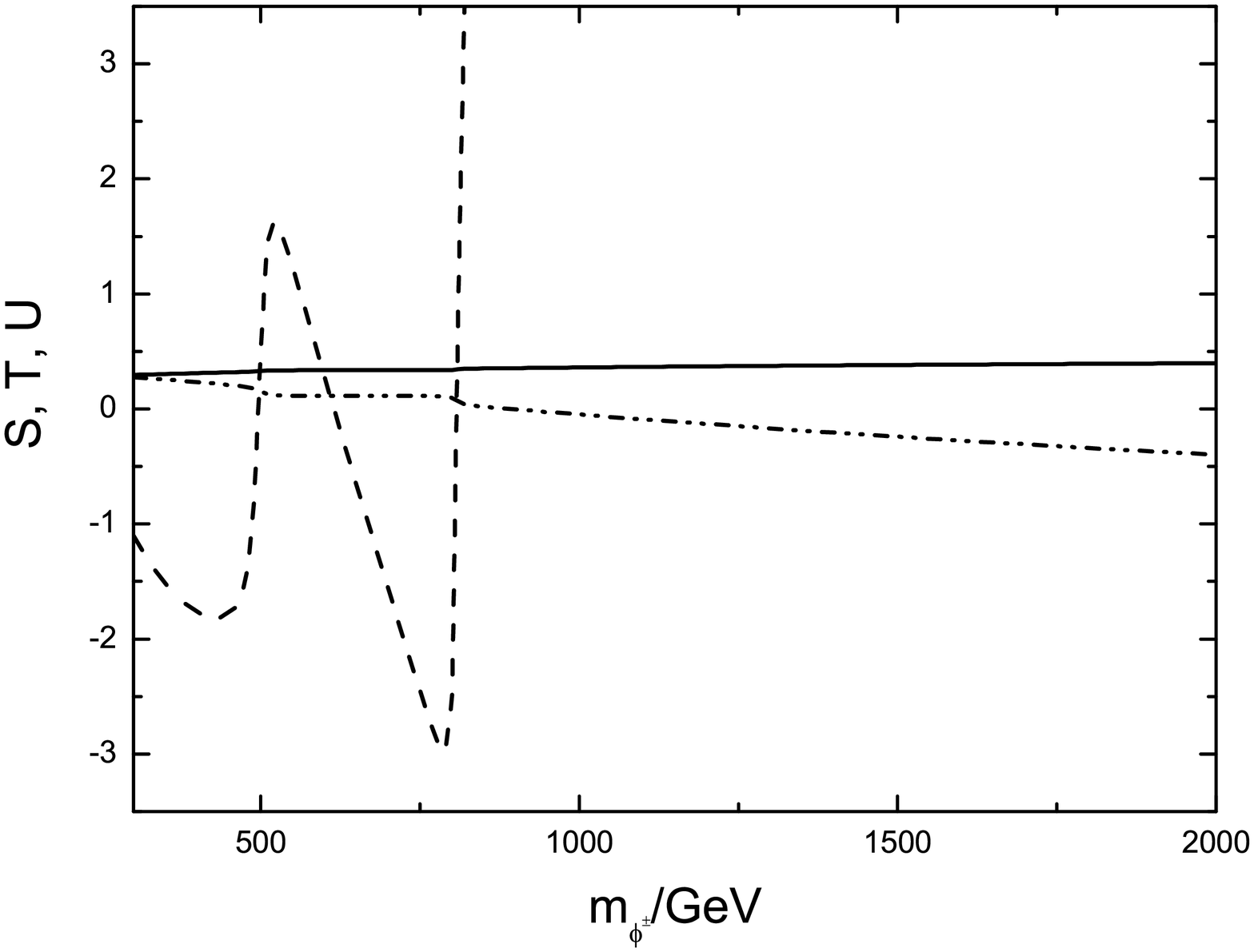}
\vspace{-1cm}
\caption[]{Adopting the assumptions mentioned in the text and assuming $\theta_1=\arg(\mu_1)=\pi,\;
\theta_2=\arg(\mu_2)=\pi/4$, we present the theoretical values for
a) $\Delta S$ (solid line), b) $\Delta U$ (dash-dot-dot line), and c) $\Delta T$ (dashed line)
versus the mass $m_{_{\phi^\pm}}$.}
\label{fig7}
\end{figure}

Choosing $\theta_1=\arg(\mu_1)=\pi,\;\theta_2=\arg(\mu_2)=\pi/4$, we depict the theoretical values of
$\Delta S,\;\Delta U,\;\Delta T$ versus the mass of charged scalar $\phi^\pm$ in Fig.(\ref{fig7}),
in which the solid line represents $\Delta S$, the dash-dot-dot line
represents $\Delta U$, and the dashed line represents $\Delta T$.
For our choices of the relevant parameters, the theoretical value of
$\Delta T$ is very sensitive to the mass $m_{_{\phi^\pm}}$, while the theoretical values of
$\Delta S and \;\Delta U$ have a weak dependence on the mass $m_{_{\phi^\pm}}$. When the mass of the charged scalar
lies in the range $400\le m_{_{\phi^\pm}}/{\rm GeV}\le 700$, the theoretical predictions of  $\Delta S,\;
\Delta T,\;\Delta U$ simultaneously satisfy the inequalities in Eq.(\ref{fitting1}).
\begin{figure}[b]
\setlength{\unitlength}{1mm}
\centering
\includegraphics[width=3.15in]{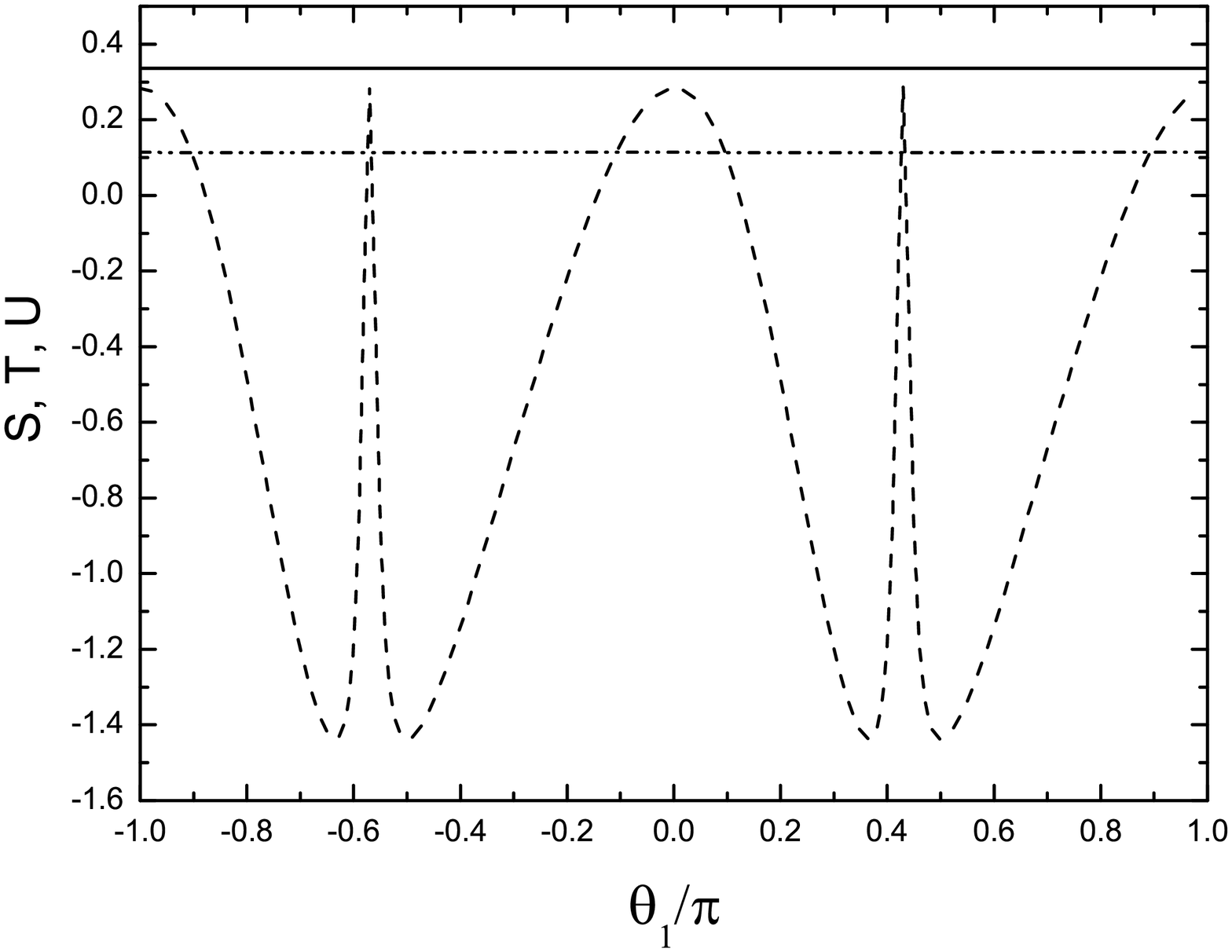}
\vspace{-1cm}
\caption[]{Adopting the assumptions mentioned in the text and assuming $m_{_{\phi^\pm}}=600\;{\rm GeV},\;
\theta_2=\arg(\mu_2)=\pi/4$, we present the theoretical values of
a) $\Delta S$ (solid line), b) $\Delta U$ (dash-dot-dot line), and c) $\Delta T$ (dashed line)
versus the CP phase $\theta_1=\arg(\mu_1)$.}
\label{fig8}
\end{figure}
The CP phases $\theta_1,\;\theta_2$ also affect the numerical results of
$\Delta S,\;\Delta U,\;\Delta T$ through the $4\times4$ mixing matrix $Z_{_{CPM}}$. Taking $m_{_{\phi^\pm}}
=600\;{\rm GeV}$ and $\theta_2=\pi/4$, we present the theoretical evaluations on
$\Delta S,\;\Delta U,\;\Delta T$ versus the CP phase $\theta_1$ in Fig.(\ref{fig8}).
With our assumptions on the parameter space, the theoretical value of
$\Delta T$ varies strongly with the CP phase $\theta_1$, while the theoretical values of
$\Delta S and \;\Delta U$ vary weakly with the CP phase $\theta_1$. In the neighbourhoods of
$\theta_1=0,\;\pm\pi/2,\;\pm\pi$, the theoretical predictions on  $\Delta S,\;
\Delta T,\;\Delta U$ simultaneously lie within the ranges presented in Eq.(\ref{fitting1}).

\begin{figure}[h]
\setlength{\unitlength}{1mm}
\centering
\includegraphics[width=3.15in]{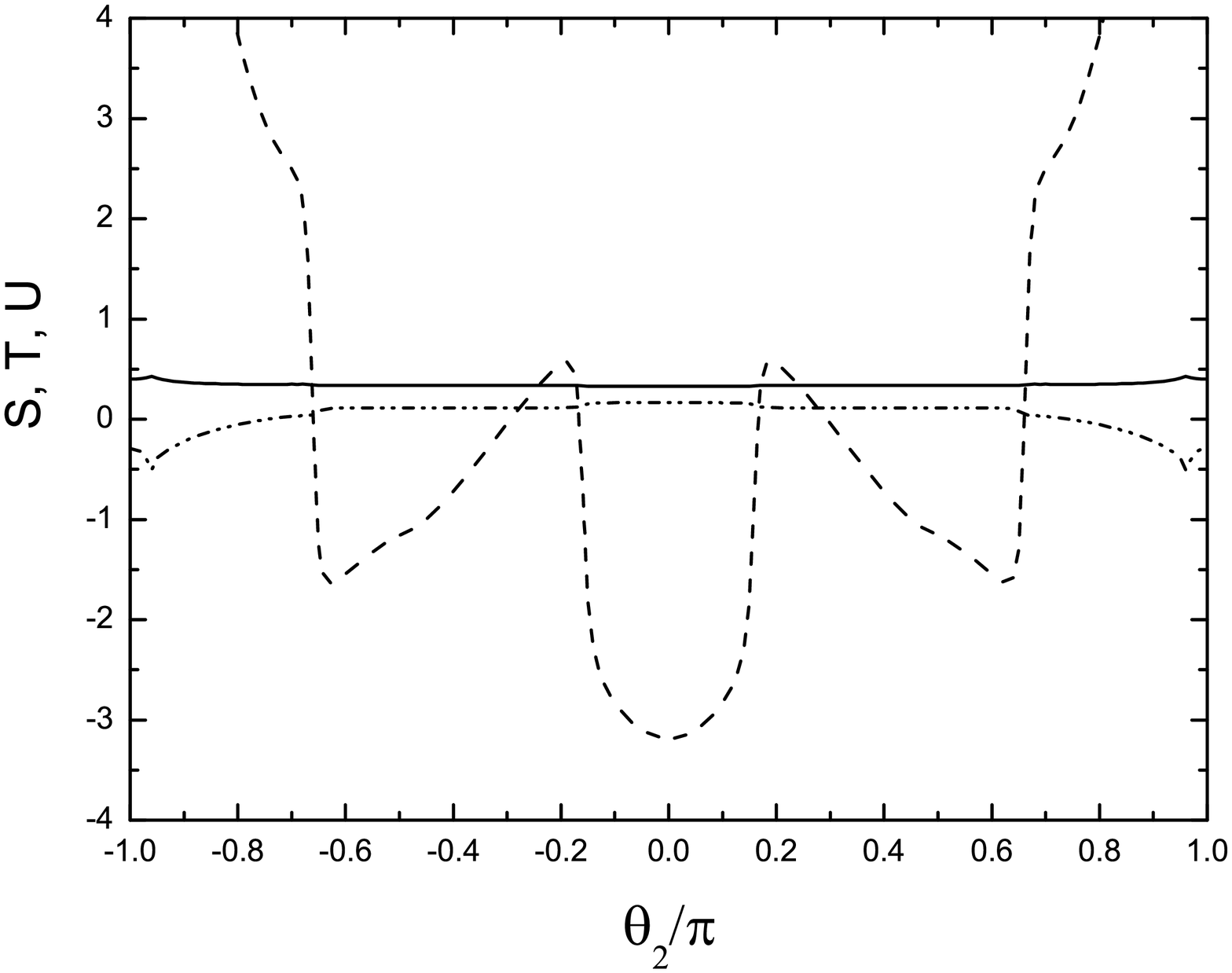}
\vspace{-1cm}
\caption[]{Adopting the mentioned assumptions in text and assuming $m_{_{\phi^\pm}}=600\;{\rm GeV},\;
\theta_1=\arg(\mu_1)=\pi$, we present the theoretical values of
a) $\Delta S$ (solid line), b) $\Delta U$ (dash-dot-dot line), and c) $\Delta T$ (dashed line)
versus the CP phase $\theta_2=\arg(\mu_2)$.}
\label{fig9}
\end{figure}

In Fig.~\ref{fig9}, we present the theoretical values of $\Delta S,\;\Delta T,\Delta U$ varying with
the CP phase $\theta_2$ when $m_{_{\phi^\pm}}=600\;{\rm GeV}$ and $\theta_1=\pi$.
As the CP phase $\theta_2$ varies, the theoretical value of $\Delta T$ changes
drastically, while the theoretical values of $\Delta S and \;\Delta U$ change slowly. In the neighbourhoods
around $\theta_2=\pm\pi/4,\;\pm3\pi/4$, the theoretical predictions of  $\Delta S,\;
\Delta T,\;\Delta U$ coincide with the present global EWPD fit within $3\sigma$ deviations.

\section{Summary\label{sec6}}
\indent\indent
For an extension of the SM with local gauged baryon and lepton numbers, we have discussed the constraints
from the oblique parameters $S,\;T,\;U$ when the lightest Higgs has a mass around $125\;{\rm GeV}$.
Considering those constraints, we find that there is parameter space to account for the excess in
Higgs production and decay in the diphoton channel observed in the ATLAS and CMS experiments.
Of course, our numerical results strongly depend on the assumptions made on the model considered here.
In other words, our theoretical prediction cannot be precise because of the theoretical uncertainties.
The purpose of our calculation is to show that this extension of the SM may be still right after the constraints
from LHC data on the Higgs and oblique parameters have been taken into account.

\appendix
\section{Higgs masses and relevant couplings\label{app1}}
\indent\indent
After diagonalizing the mass matrix Eq.(\ref{mass-matrix1}), we obtain
\begin{eqnarray}
&&m_{_{h_0}}^2={\it Min}(m_1^2,\;m_2^2,\;m_3^2)
\nonumber\\
&&m_{_{H_3^0}}^2={\it Max}(m_1^2,\;m_2^2,\;m_3^2)
\label{Higgs-mass3}
\end{eqnarray}
with
\begin{eqnarray}
&&m_1^2=-{a\over3}+{2\over3}p\cos\phi\;,
\nonumber\\
&&m_2^2=-{a\over3}-{1\over3}p(\cos\phi-\sqrt{3}\sin\phi)\;,
\nonumber\\
&&m_3^2= -{a\over3}-{1\over3}p(\cos\phi+\sqrt{3}\sin\phi)\;.
\label{Higgs-mass4}
\end{eqnarray}

To formulate the expressions in a concise form, we define the notations
\begin{eqnarray}
&&p=\sqrt{a^2-3b},\;\phi={1\over3}\arccos(-{1\over p^3}(a^3-{9\over2}ab+{27\over2}c))
\label{Higgs-mass5}
\end{eqnarray}
where
\begin{eqnarray}
&&a=-2\Big(\lambda_{_{HH}}\upsilon^2+\lambda_{_{BB}}\upsilon_{_B}^2+\lambda_{_{LL}}\upsilon_{_L}^2\Big)\;,
\nonumber\\
&&b=4(\lambda_{_{HH}}\lambda_{_{BB}}\upsilon^2\upsilon_{_B}^2+\lambda_{_{HH}}\lambda_{_{LL}}\upsilon^2\upsilon_{_L}^2
+\lambda_{_{BB}}\lambda_{_{LL}}\upsilon_{_B}^2\upsilon_{_L}^2)
\nonumber\\
&&\hspace{0.8cm}
-\lambda_{_{HB}}^2\upsilon^2\upsilon_{_B}^2-\lambda_{_{HL}}^2\upsilon^2\upsilon_{_L}^2
-\lambda_{_{BL}}^2\upsilon_{_B}^2\upsilon_{_L}^2\;,
\nonumber\\
&&c=2\Big(\lambda_{_{HH}}\lambda_{_{BL}}^2+\lambda_{_{BB}}\lambda_{_{HL}}^2
+\lambda_{_{LL}}\lambda_{_{HB}}^2-4\lambda_{_{HH}}\lambda_{_{BB}}\lambda_{_{LL}}
\nonumber\\
&&\hspace{0.8cm}
-\lambda_{_{HB}}\lambda_{_{HL}}\lambda_{_{BL}}\Big)\upsilon^2\upsilon_{_B}^2\upsilon_{_L}^2\;.
\label{Higgs-mass6}
\end{eqnarray}
The normalized eigenvectors of the mass squared matrix in Eq.(\ref{mass-matrix1}) are given by
\begin{eqnarray}
&&\left(\begin{array}{c}\Big(Z_{_{CPE}}\Big)_{11}\\
\Big(Z_{_{CPE}}\Big)_{21}\\\Big(Z_{_{CPE}}\Big)_{31}
\end{array}\right)={1\over\sqrt{|X_1|^2+|Y_1|^2+|Z_1|^2}}\left(\begin{array}{c}
X_1\\Y_1\\Z_1\end{array}\right)\;,
\nonumber\\
&&\left(\begin{array}{c}\Big(Z_{_{CPE}}\Big)_{12}\\
\Big(Z_{_{CPE}}\Big)_{22}\\\Big(Z_{_{CPE}}\Big)_{32}
\end{array}\right)={1\over\sqrt{|X_2|^2+|Y_2|^2+|Z_2|^2}}\left(\begin{array}{c}
X_2\\Y_2\\Z_2\end{array}\right)\;,
\nonumber\\
&&\left(\begin{array}{c}\Big(Z_{_{CPE}}\Big)_{13}\\
\Big(Z_{_{CPE}}\Big)_{23}\\\Big(Z_{_{CPE}}\Big)_{33}
\end{array}\right)={1\over\sqrt{|X_3|^2+|Y_3|^2+|Z_3|^2}}\left(\begin{array}{c}
X_3\\Y_3\\Z_3\end{array}\right)\;,
\label{Higgs-mass7}
\end{eqnarray}
with
\begin{eqnarray}
&&X_1=(2\lambda_{_{BB}}\upsilon_{_B}^2-m_1^2)(2\lambda_{_{LL}}\upsilon_{_L}^2-m_1^2)
-\lambda_{_{BL}}^2\upsilon_{_B}^2\upsilon_{_L}^2\;,
\nonumber\\
&&Y_1=\lambda_{_{HL}}\lambda_{_{BL}}\upsilon\upsilon_{_B}\upsilon_{_L}^2-\lambda_{_{HB}}\upsilon\upsilon_{_B}
(2\lambda_{_{LL}}\upsilon_{_L}^2-m_1^2)\;,
\nonumber\\
&&Z_1=\lambda_{_{HB}}\lambda_{_{BL}}\upsilon\upsilon_{_B}^2\upsilon_{_L}-\lambda_{_{HL}}\upsilon\upsilon_{_L}
(2\lambda_{_{BB}}\upsilon_{_B}^2-m_1^2)\;,
\nonumber\\
&&X_2=\lambda_{_{HL}}\lambda_{_{BL}}\upsilon\upsilon_{_B}\upsilon_{_L}^2-\lambda_{_{HB}}\upsilon\upsilon_{_B}
(2\lambda_{_{LL}}\upsilon_{_L}^2-m_2^2)\;,
\nonumber\\
&&Y_2=(2\lambda_{_{HH}}\upsilon^2-m_2^2)(2\lambda_{_{LL}}\upsilon_{_L}^2-m_2^2)
-\lambda_{_{HL}}^2\upsilon^2\upsilon_{_L}^2\;,
\nonumber\\
&&Z_2=\lambda_{_{HB}}\lambda_{_{HL}}\upsilon^2\upsilon_{_B}\upsilon_{_L}-\lambda_{_{BL}}\upsilon_{_B}\upsilon_{_L}
(2\lambda_{_{HH}}\upsilon^2-m_2^2)\;,
\nonumber\\
&&X_3=\lambda_{_{HB}}\lambda_{_{BL}}\upsilon\upsilon_{_B}^2\upsilon_{_L}-\lambda_{_{HL}}\upsilon\upsilon_{_L}
(2\lambda_{_{BB}}\upsilon_{_B}^2-m_3^2)\;,
\nonumber\\
&&Y_3=\lambda_{_{HB}}\lambda_{_{HL}}\upsilon^2\upsilon_{_B}\upsilon_{_L}-\lambda_{_{BL}}\upsilon_{_B}\upsilon_{_L}
(2\lambda_{_{HH}}\upsilon^2-m_3^2)\;,
\nonumber\\
&&Z_3=(2\lambda_{_{HH}}\upsilon^2-m_3^2)(2\lambda_{_{BB}}\upsilon_{_B}^2-m_3^2)
-\lambda_{_{HB}}^2\upsilon^2\upsilon_{_B}^2\;.
\label{Higgs-mass8}
\end{eqnarray}

\section{The loop functions\label{app2}}
\indent\indent
The loop functions in Eq.(\ref{hgg}) and Eq.(\ref{hpp}) are given as
\begin{eqnarray}
&&A_1(x)=-\Big[2x^2+3x+3(2x-1)g(x)\Big]/x^2\;,\nonumber\\
&&A_{1/2}(x)=2\Big[x+(x-1)g(x)\Big]/x^2\;,\nonumber\\
&&A_0(x)=-(x-g(x))/x^2\;,
\label{loop-functions}
\end{eqnarray}
with
\begin{eqnarray}
&&g(x)=\left\{\begin{array}{l}\arcsin^2\sqrt{x},\;x\le1\\
-{1\over4}\Big[\ln{1+\sqrt{1-1/x}\over1-\sqrt{1-1/x}}-i\pi\Big]^2,\;x>1\end{array}\right.
\label{g-function}
\end{eqnarray}

\end{document}